\documentclass[useAMS,usenatbib,mnras]{mn2e}
\usepackage{aas_macros}
\usepackage[pdftex]{graphicx}
\usepackage{amsmath}
\usepackage{deluxetable}

\def\pasa{Publ. Astron. Soc. Australia}
\def\psra{PSR~J1705$-$3950}
\def\psrb{PSR~J1638$-$4608}

\def\psrf{PSR~B1828$-$11}

\title[Periodic Signals from Young Pulsars]{Periodic modulation in pulse arrival times from young pulsars: a renewed case for neutron star precession}

\author[M. Kerr, G. Hobbs, S. Johnston, R. M. Shannon]
{ M.~Kerr$^{1}$\thanks{E-mail: matthew.kerr@gmail.com},
  G.~Hobbs$^{1}$,
  S.~Johnston$^{1}$,
  and R.~M.~Shannon$^{1}$ \\
$^{1}$CSIRO Astronomy and Space Science, Australia Telescope National
Facility, PO~Box~76, Epping NSW~1710, Australia}
\begin{document}

\date{Accepted 2015 October 20. Received 2015 October 16; in original form
2015 September 18}

\pagerange{\pageref{firstpage}--\pageref{lastpage}} \pubyear{2015}

\maketitle

\label{firstpage}

\begin{abstract}
In a search for periodic variation in the arrival times of pulses from
151 young, energetic pulsars, we have identified seven cases of
modulation consistent with one or two harmonics of a single
fundamental with time-scale 0.5--1.5\,yr.  We use simulations to show
that these modulations are statistically significant and of high
quality (sinusoidal) even when contaminated by the strong stochastic
timing noise common to young pulsars.  Although planetary companions
could induce such modulation, the large implied masses and 2:1 mean
motion resonances challenge such an explanation.  Instead, the
modulation is likely to be intrinsic to the pulsar, arising from
quasi-periodic switching between stable magnetospheric states, and we
propose that precession of the neutron star may regulate this
switching.
\end{abstract}


\section{Introduction}

In the \textit{tour de force} publication announcing the discovery of
pulsars, \citet{Hewish68} made the first use of periodic modulation of
the pulse time of arrival (TOA), induced by the earth's motion around
the sun, to pinpoint the position of PSR~B1919$+$21.  The study of TOA
modulation has ever since been the heart of pulsar science.  After
astrometry, the most common manifestation of modulation arises from
the reflex motion induced by a companion, notably leading to the
discoveries of the relativistic double neutron star system B1913$+$16
\citep{Hulse75} and the first exoplanets orbiting PSR~B1257+12 \citep{Wolszczan90}.

Later, \citet{Stairs00} reported quasi-periodic modulation of the TOAs
of PSR~B1828$-$11 (PSR~J1830$-$1059) which was correlated with changes
in the pulse shape and proposed free precession of the neutron star as
a natural explanation of these observations.  In brief, the precession
induces a periodic change in spin-down torque, leading to TOA
modulation, and the precessional motion allows the observer to see
different portions of the neutron star polar cap, gradually changing
the pulse emission profile.  Various authors explored the implications
for pulsar beam shapes and geometry \citep[e.g.][]{Jones01,Link01} and
for neutron star interiors \citep[e.g.][]{Cutler03,Link03}.

A decade later, \citet{Lyne10} identified similar quasi-periodic
modulation in 17 pulsars drawn from a sample of 366 pulsars
\citep{Hobbs10} timed over
decades.  Six of these pulsars showed correlated profile variation;
however, they also observed the pulse profile to oscillate between two
distinct shapes on short time-scales.  This behaviour suggests that the
periodicity is not due to secular motion of the neutron star but to
rapid reconfiguration of the pulsar magnetosphere between two
spin-down states.  \citet{Lyne10} further argued that state switching
with time-scales randomly distributed around a fundamental value could
fully explain the observed TOA modulation.  Consequently, the
precession hypothesis fell out of favour.  However, the year-long
time-scale underpinning the state switching is difficult to produce
from any dynamic process in the pulsar magnetosphere
\citep[e.g.][]{Cordes13}.

A better understanding of both the magnetospheric switching process
and its regularity is hampered by the paucity of such pulsars; even
the large parent sample \citet{Lyne10} produced only 17 examples of
TOA modulation, and most of these cases are not of high quality
factor.  Here, we report the discovery of nearly sinusoidal TOA
modulation from seven pulsars drawn from a sample of 151 young,
energetic (spin-down power $\dot{E}>10^{34}$\,erg\,s$^{-1}$) pulsars.
The short spin periods are complementary to the sample of
\citet{Lyne10} and offer the opportunity to study the modulation
period $P_m$ over a broad range of pulsar ages and spin periods $P_s$.
Combined measurements of $P_m$ from other neutron stars
\citep{Jones12}, we find a strong correlation between 
$P_m$ and $P_s$, a scaling that can be readily explained if neutron
star precession drives the ``switching clock''.  However, as we
discuss below, many difficulties with such a mechanism remain.

Below, we outline our search for periodic modulations and the resulting
candidates (\S\ref{sec:data}).  The identification of these signals is
complicated by ``timing noise'', and in \S\ref{sec:cands} we detail
the simulations we perform to verify and characterise the candidates.
In \S\ref{sec:profile}, we search for variations in the pulse profile
of our candidate but find no strong evidence for them.  Finally, we
discuss in \S\ref{sec:discussion} the possible causes of the TOA
modulation, including planetary companions and a hybrid scenario of
magnetospheric switching with free precession.

\section{Search for Periodic Signals}
\label{sec:data}

Our sample of 151 pulsars, the data (raw and reduced), the maximum
likelihood method of analysis, and the search for periodic signals (in
the context of upper limits on planetary companions) are described
fully in \citet{Kerr15}.  For completeness, we summarise this
discussion and review the search for quasi-periodic signals.

The TOAs are modelled as a parametric ``spin-down'' process which we
evaluate with \textsc{tempo2} \citep{Hobbs06}.  To the residuals of
this model are added components of white noise---both measurement
(radiometer) and ``jitter'' noise \citep{Shannon14})---and strong, red
``timing'' noise which we model as a widesense stationary process
whose spectrum is described by a power law with a low-frequency cutoff
\citep{Coles11}:
\begin{equation}
\label{eq:noise}
P(f) = A_0 \left[1 + (f/f_c)^2\right]^{-\alpha/2}.
\end{equation}
These noise components manifest as the diagonal and non-diagonal
components of a covariance matrix for the residuals, $C$, which allows
evaluation of the Gaussian log likelihood for the full (spin-down plus
noise) model.  We use Monte Carlo Markov Chains methods, in particular
the \texttt{emcee} package \citep{Foreman-Mackey13}, to evaluate the
likelihood and characterize its shape, allowing parameter inference.

To search for harmonic modulation, we simply add a sinusoid with
arbitrary amplitude and phase to the spin-down model and compare the
change in log likelihood $\delta\log\mathcal{L}$ between the two
best-fitting models.  The significance of a given value of
$\delta\log\mathcal{L}$ will generally depend on the particular noise
characteristics of a pulsar, but metrics like the Akaike information
criterion suggest that values of $\delta\log\mathcal{L}>6$ are
significant for three degrees of freedom.  Because we also have a
sample of 151 pulsars, or ``trials'',  we adopt
$\delta\log\mathcal{L}>10$, for any trial harmonic frequency, as a
threshold for further consideration of a pulsar as a modulation
candidate.

\begin{figure}
\includegraphics[angle=0,width=0.45\textwidth]{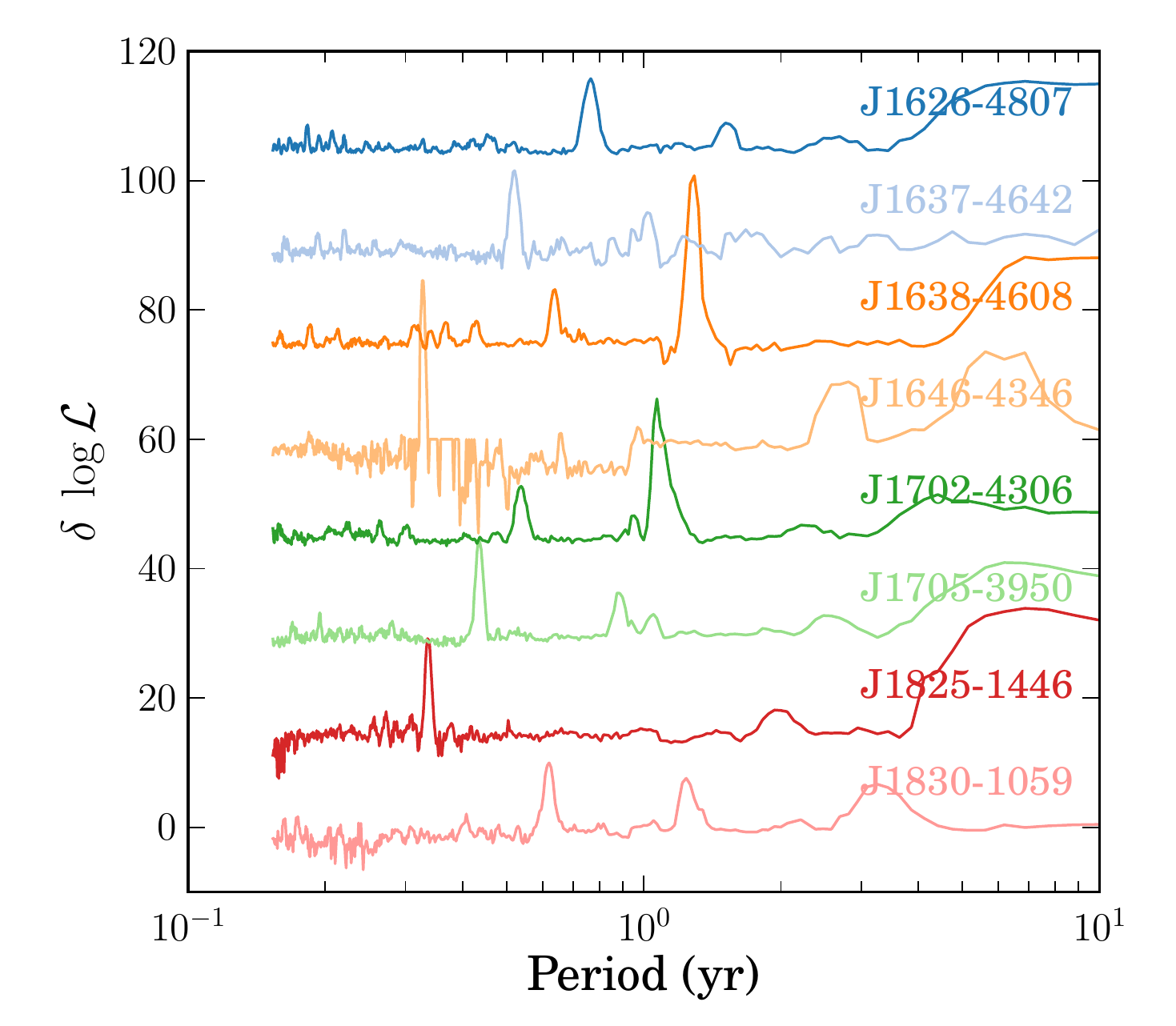}
\caption{\label{fig:ts_plot}The change in log likelihood ($\delta
\log\mathcal{L}$) between a model with and without a single sinusoidal
modulation at the indicated period, with an arbitrary vertical offset of
$\delta\log\mathcal{L}=15$ between pulsars for clarity.}
\end{figure}

\begin{figure}
\includegraphics[angle=0,width=0.45\textwidth]{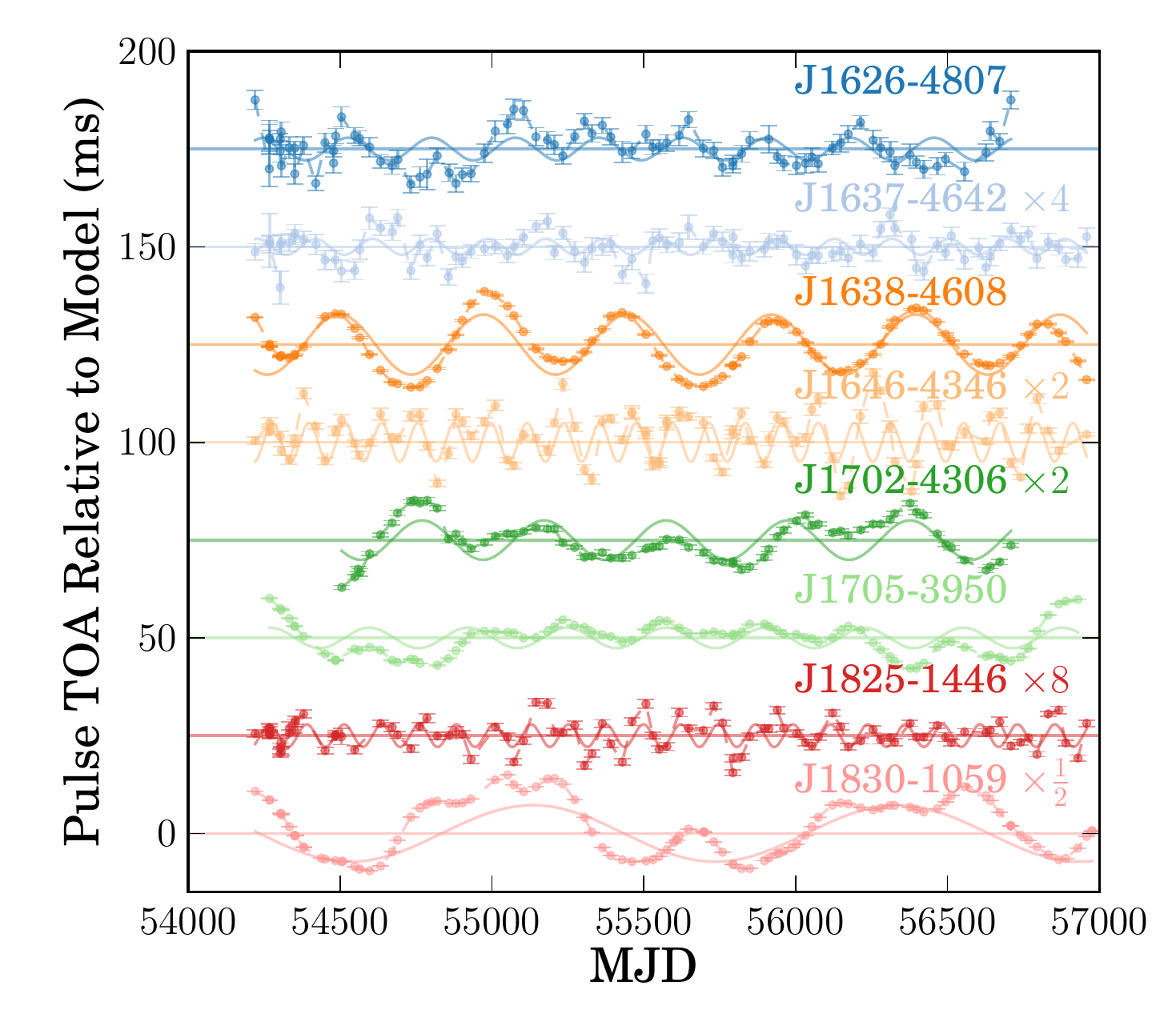}
\caption{\label{fig:residuals}The deviation in arrival times of pulses
from the indicated pulsars relative to a simple timing model including
a cubic term. Due to the higher level of red noise, we have
additionally removed a series of six sinusoids (PSRs~J1637$-$4642 and
J1825$-$1446) and eight sinusoids (PSR~J1646$-$4346).  We note that
these frequencies are substantially lower than the detected modulation
frequencies.  For clarity, the residuals are connected by a dashed line
and have been multiplied by the indicated factor to ease comparison.
The best-fit single-sinusoid model is indicated by the solid line.}
\end{figure}

The $\delta\log\mathcal{L}$ for the eight pulsars surpassing
this threshold, as a function of trial harmonic modulation period, appears
in Figure \ref{fig:ts_plot}, which can be thought of as a periodogram
that properly accounts for the red noise.  In such a periodogram,
representation of a pure sinusoid as a Dirac $\delta$ function is
modified by the ``window function'', determined by the length,
cadence, and noise properties of the data for a given pulsar.  The
window function is not known \textit{a priori}, but with the exception
of \psrf{}, the widths of the peaks are generally consistent with
signals caused by sinusoidal modulation.  We verify this below with
simulations.  A clear second harmonic, whose significance and nature
we also explore below, is present in some pulsars.

The timing residuals to simple spin-down models that remove low
frequency timing noise appear in Figure \ref{fig:residuals}.  The
quasi-periodic oscillation is evident to the eye in most.
Anecdotally, our threshold of $\delta\log\mathcal{L}>10$ seems to
reproduce what the eye picks up in the time domain.  It seems likely,
then, that there are additional pulsars harbouring such signals that
fail to rise above the noise in our analysis.  Moreover, pulsars like
PSR~B1828$-$11, with multiple time-scales, are not well modelled by
our single-frequency test, yielding lower significances.  However,
searches for multiple sinusoids are computationally prohibitive, and
we have visually examined the timing residuals for all 151 pulsars to
search for any additional ``obvious'' modulation, finding none.  As
noted by \citet{Kerr15b}, the presence of strong timing noise causes
sensitivity to coherent sinusoids to scale roughly as the square root
of observing span. Thus, for most of our sample, substantial further
observations will be required to detect modulations with lower
amplitude.

\section{Characterisation of periodic signals}
\label{sec:cands}

Characterising periodicity in astronomical signals is a common yet
often challenging task.  One might, e.g., search for peaks in the time
series power spectrum, but nonuniform observing cadence and
sensitivity (e.g.  due to changes in source brightness) hamper simple
interpretations of spectral features.  This difficulty is encapsulated
in the ``window function'' and the well-known convolution theorem.
Observing with nonuniform cadence and sensitivity is equivalent to
multiplying the true signal by a complicated weighting window
function.  The simplest window function, a top hat, is Fourier
transformed into a sinc; thus, the power spectrum of a finite-length
time series will be convolved with a function that diminishes
asymptotically as $f^{-2}$.  This effect, which blurs spectral
features and, for steep spectra, shifts low frequency power to high
frequencies, is known as spectral leakage.

Our data have nonuniform cadence and hetereogeneous sensitivity (due
to system evolution and varying source brightness) and are affected by
timing noise with a steep red spectrum.  While we can mitigate
spectral leakage following the methods of \citet{Coles11}, the
quasi-sinusoidal signals in the data are sufficiently strong that they
bias the measurement of the timing noise parameters.  These
considerations rule out spectral analysis as a primary method for
identifying and characterising modulations in our data; instead, we
outline below a prescription of simulations to cleanly separate the
effects of timing noise and sinusoidal modulation.  We conclude the
section with a discussion of two pulsars, B1828$-$11 and B0740$-$28,
whose spectral features indicate strong modulation but which our
analysis is able to show are of lower quality.

\subsection{Simulations}

To quantify both the significance and the quality factor of the
periodic modulation, we generated a series of simulated realizations
of our TOAs.  For our control case (S0), we adopted a red-noise model
similar to that observed in the data after removing the apparent
sinusoidal modulation.  We next generated two additional sets of
simulations with the same red-noise model plus a single sinusoid (S1)
and two sinusoids (S2); we only performed S2 for those pulsars with
evidence of a harmonic.  The observed and simulated parameters appear
in Table \ref{tab:parms}.  In the table, we do not report the measured
values of the spin noise parameters as they are typically covariant.
Instead, we have chosen representative values that both yield the
correct value for the likelihood (discussed below) and reproduce the
observed spectrum of the time series residuals.  To form each of 100
realizations, we used the original TOA sampling and a simple timing
solution to generate an ``idealized'' set of TOAs.  To these we added
white noise from measurement error and ``jitter'', the latter at the
values determined from a model of the data.  To these white-noise
realizations, we added a random time series with a spectrum matching
the red-noise model.

\begin{figure}
\includegraphics[angle=0,width=0.45\textwidth]{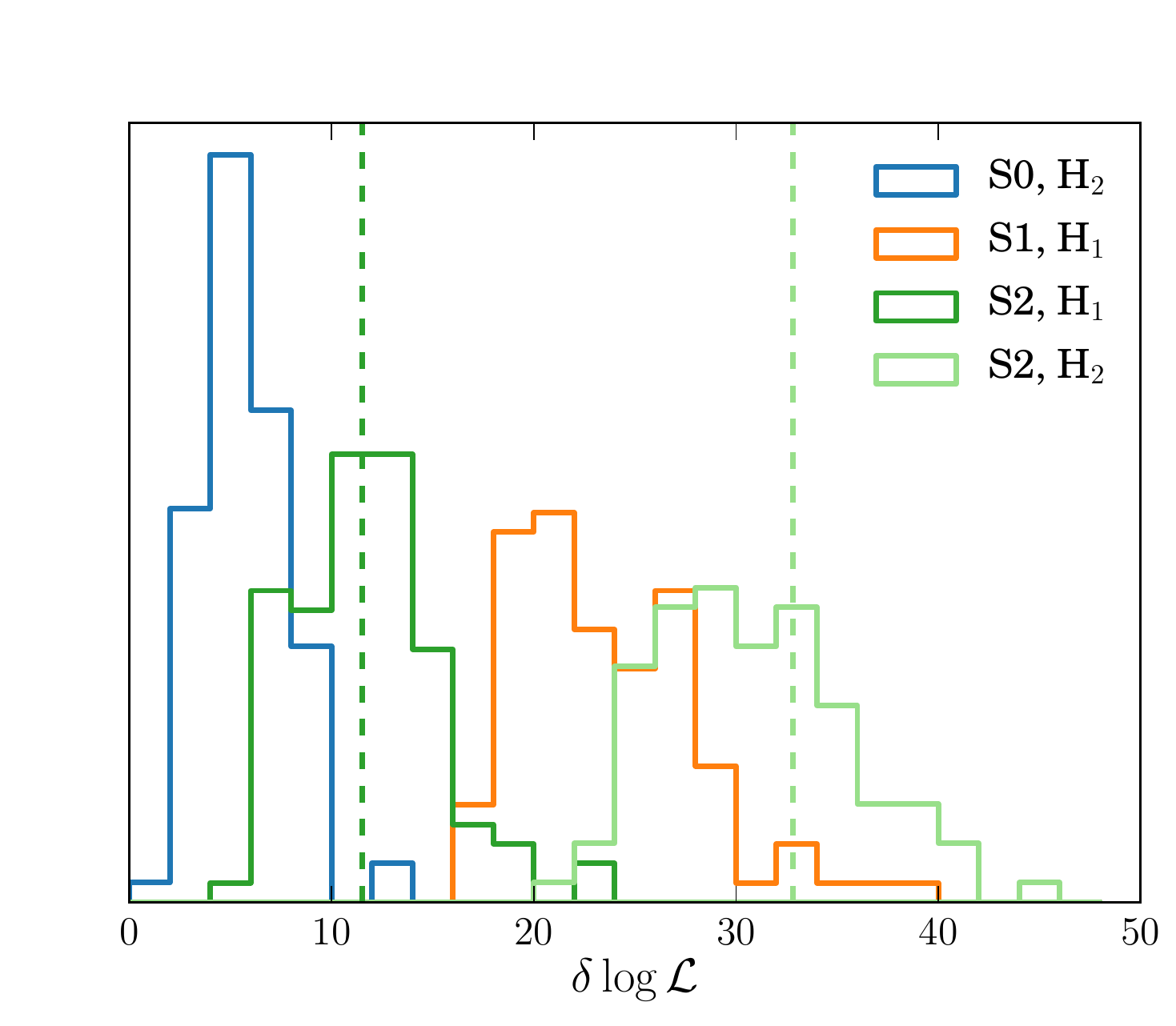}
\caption{\label{fig:j1705_hist}Distributions of the change in $\log$
likelhihood for the simulations of PSR~J1705$-$3950 described in the
main text; all values are relative to $\log\mathcal{L}(H_0)$ for the
indicated simulation.  The observed values
$\delta\log\mathcal{L}=11.5$ and $\delta\log\mathcal{L}=32.8$ are
shown with vertical dashed lines.}
\end{figure}

For each set of simulations, we followed the same method as for the
real data and determined best-fit parameters under each of the three
hypotheses: $H_0$, no sinusoids; $H_1$, a single sinusoid; and $H_2$,
two sinusoids.  We then compared the change in the log likelihood,
$\delta\log\mathcal{L}$, as a measure of both significance and
goodness of fit.  The permutations of $S$ and $H$ allow a variety of
statistical tests.  E.g., the distribution of
$\log\mathcal{L}(H_2)-\log\mathcal{L}(H_0)$ using data from S0
allows us to test how often noise fluctuations produce large values of
$\delta\log\mathcal{L}$ by chance when we search for two sinusoids.

As an example, we consider the pulsar with the strongest and most
complex modulation, PSR~J1705$-$3950, which has evidence for a second
harmonic.  The distributions of $\delta\log\mathcal{L}$ for some of
the permutations of $S$ and $H$ appear in Figure \ref{fig:j1705_hist}.
The first distribution (blue in the Figure, leftmost) is a
significance test, in which we simulate only red noise (S0) but fit a
model with two sinusoids ($H_2$); its properties are reported in the
rightmost column of Table \ref{tab:logls}.  The realizations of
$\delta\log\mathcal{L}$ are all well below the observed value under
$H_2$, meaning the modulations cannot be produced by noise
fluctuations.

The next distribution to the right (green) contains the results for a
simulation with two sinusoids (S2) while only one is modelled ($H_1$);
the values correspond to the middle column and last two rows of the
relevant entry in Table \ref{tab:logls}.  This combination tests the
consistency of the two-sinusoid model by examining the significance we
achieve when we only model part of the modulation, and we see that the
observed value of $\delta\log\mathcal{L}(H_1)=11.5$ lies within the
core of the distribution of simulated values, implying consistency of
the two-sinusoid simulations with the data.

Next is the distribution for a simulation with only one sinusoid (S1) and the corresponding
change in log likelihood under H$_1$.  This distribution is in strong
disagreement with the observed value of
$\delta\log\mathcal{L}(H_1)$=11.5.  For simplicity of presentation, we
have excluded the distribution of
$\delta\log\mathcal{L}(H_2)-\delta\log\mathcal{L}(H_1)$, which
provides the best significance test of the harmonic, but these results
are tabulated in the rightmost column of S1 for the relevant
pulsars.

Finally, rightmost (light green) is the distribution of
$\delta\log\mathcal{L}$ for the scenario we believe describes the data
best, viz. two sinusoids.  The observed value again lies within the
core of the simulated values.  This is a key point, both in terms of
the significance of the model and in goodness of fit.  Not only does
the two-sinusoid hypothesis (H$_2$) produce a $\delta\log\mathcal{L}$
much larger than that produced by chance, but the observed value of
$\delta\log\mathcal{L}$ is precisely the same as that observed for the
simulations.  That is, the data are completely consistent with high
quality factor modulations.  This test is equivalent to, and arguably
better than, jackknife tests where we search for the signal at lower
significance in subsets of the data.  Perhaps most interestingly, the
\textit{absolute} value of $\log\mathcal{L}$ agrees between the
simulations and data.  This means that our model completely captures
all of the information in the data, such that our assumptions about
the noise model, spin-down model, and modulations are correct and
complete.

PSR~J1705$-$3950 is our best example; its low white and red noise,
together the highest ratio of harmonic to fundamental amplitudes in the
sample, make it ideal for detecting all of the features.  But similar
remarks apply to other candidates in various levels of dilution: most
lack evidence for a second harmonic, but the data are perfectly
consistent with and statistically favour at least one sinusoid.  The
full results of the simulations are compiled in Table \ref{tab:logls}.
In summary, they are: PSRs~J1626$-$4807 and J1637$-$4642 show marginal
evidence for a single sinusoid; PSRs J1646$-$4346 and J1825$-$1446
show strong evidence for a single sinusoid; PSRs J1638$-$4608 and
J1702$-$4306 show strong evidence for a single sinusoid and modest
evidence for a harmonic; and PSR~J1705$-$3950 shows strong evidence
for a fundamental and a harmonic.

\begin{table*}
\caption{\label{tab:parms} Observed and simulated parameters.  The
stochastic spin noise parameters are those of Equation \ref{eq:noise},
and the modulation parameters are reported as Keplerian parameters,
specifically the period $P_{bn}$, epoch of periastron $T_{n}$, and
projected semi-major axis $a_n$ of the $n$th orbiting body.  All
simulated noise models have $f_c=0.05$\,yr$^{-1}$.}
\begin{tabular}{lllllllll}
\hline
& $A_0$ & $\alpha$ & $a_1$ & $P_{b1}$ & $T_{1}$ & $a_2$ & $P_{b2} $ & $T_2$ \\
& yr$^{-3}$ &  & ms & d & MJD & ms & d & MJD \\
\hline
PSR~J1626$-$4807 & -- & -- & 2.9(7) & 276(5) & 55560(10) &
-- & -- & \\
\phantom{PSR~}simulation & $7.0\times10^{-17}$ &3.5 &  2.9 &
276 & 55560 & -- & --& -- \\
\hline
PSR~J1637$-$4642 & -- & -- & 0.5(1) & 190(2) & 55510(7) &
-- & -- & \\
\phantom{PSR~}simulation & $2.0\times10^{-13}$ & 6.2 &  0.5 &
190 & 55510 & -- & --& -- \\
\hline
PSR~J1638$-$4608 & -- & -- & 8.4(10) & 470(5) & 55327(8) & 0.78(18) & 236(3) & 55374(9) \\
\phantom{PSR~}simulation & $1.4\times10^{-15}$ & 4.5 &  8.4 &
470 & 55328 & 0.78 & 236 & 55374 \\
\hline
PSR~J1646$-$4346 & -- & -- & 2.9(7) & 276(5) & 55560(10) &
-- & -- & --\\
\phantom{PSR~}simulation & $5.8\times10^{-14}$ & 4.5 &  2.5 &
120 & 55565 & -- & --& -- \\
\hline
PSR~J1702$-$4306 & -- & -- & 2.9(6) & 391(10) & 55473(7) &
0.42(8) & 198(2) & 55541(6) \\
\phantom{PSR~}simulation & $2.1\times10^{-17}$ & 4.0 &  2.9 & 391
& 55473 & 0.42 & 198 & 55541\\
\hline
PSR~J1705$-$3950 & -- & -- & 2.6(3) & 326(4) & 55495(7) & 0.67(9) & 159.5(8) & 55507 (4) \\
\phantom{PSR~}simulation & $5.5\times10^{-18}$ & 3.0 &  2.5 & 326 & 55494 & 0.69 & 159.5 & 55507 \\
\hline
PSR~J1825$-$1446 & -- & -- & 0.36(7) & 123(1) & 55468(4) & -- &
-- & -- \\
\phantom{PSR~}simulation & $3.6\times10^{-17}$ & 3.5 &  0.36 &
123 & 55468 & -- & --& -- \\
\end{tabular}
\end{table*}

\begin{table}
\caption{\label{tab:logls} Log likelihod values from simulations and
data.  The figures give the median values, and the upper and lower
error bars the central 68\% containment interval.}
\begin{tabular}{lrrr}
\hline
 &   $\log\mathcal{L}(H_0)$  &  $\log\mathcal{L}(H_1)-$& $\log\mathcal{L}(H_2)-$ \\ 
 &                           &  $\log\mathcal{L}(H_0)$\phantom{$-$} & $\log\mathcal{L}(H_0)$\phantom{$-$}\\ 
 \hline
PSR~J1626$-$4807 \\
S0 &   $454.8_{-9.3}^{+6.0}$  &  $5.8_{-1.5}^{+1.9}$  &  --  \\ 
S1 &   $445.0_{-8.3}^{+5.5}$  &  $11.8_{-2.6}^{+2.3}$  &  --  \\ 
Data  & 440.3\phantom{$_{-0.0}^{+0.0}$} &
9.9\phantom{$_{-0.0}^{+0.0}$} &  -- \\
\hline
PSR~J1637$-$4642 \\
S0 &   $556.6_{-6.3}^{+10.5}$  &  $4.2_{-1.4}^{+2.0}$  &  --  \\ 
S1 &   $552.2_{-11.2}^{+8.4}$  &  $7.7_{-2.8}^{+5.4}$  &  --  \\ 
Data  & 551.5\phantom{$_{-0.00}^{+0.00}$} &
11.7\phantom{$_{-0.0}^{+0.0}$} & -- \\
\hline
PSR~J1638$-$4608 \\
S0 &   $593.8_{-7.2}^{+7.4}$  &  $8.0_{-3.1}^{+3.9}$  &  $9.0_{-2.7}^{+3.5}$  \\ 
S1 &   $578.5_{-5.2}^{+8.0}$  &  $16.0_{-3.2}^{+3.5}$  &  $19.4_{-3.6}^{+3.9}$  \\ 
S2 &   $575.8_{-5.8}^{+7.6}$  &  $12.2_{-2.4}^{+3.2}$  &  $20.5_{-3.6}^{+3.8}$  \\ 
Data  & 576.3\phantom{$_{-0.0}^{+0.0}$} & 14.0\phantom{$_{-0.0}^{+0.0}$} & 23.9\phantom{$_{-0.0}^{+0.0}$} \\
\hline
PSR~J1646$-$4346 \\
S0 &   $493.5_{-5.1}^{+6.8}$  &  $3.4_{-1.3}^{+1.8}$  &  --  \\ 
S1 &   $470.2_{-8.3}^{+6.2}$  &  $25.0_{-5.1}^{+5.6}$  &  --  \\ 
Data  & 471.1\phantom{$_{-0.0}^{+0.0}$} &
26.3\phantom{$_{-0.0}^{+0.0}$} & -- \\
\hline
PSR~J1702$-$4306 \\
S0 &   $510.0_{-8.5}^{+5.9}$  &  $3.2_{-2.4}^{+2.5}$  &  $5.3_{-1.5}^{+1.8}$  \\ 
S1 &   $494.1_{-5.7}^{+7.4}$  &  $17.4_{-2.8}^{+4.2}$  &  $20.7_{-3.9}^{+4.7}$  \\ 
S2 &   $489.8_{-6.2}^{+7.1}$  &  $11.1_{-2.8}^{+1.7}$  &  $23.5_{-2.8}^{+5.7}$  \\ 
Data  & 485.5\phantom{$_{-0.0}^{+0.0}$} & 10.0\phantom{$_{-0.0}^{+0.0}$} & 23.5\phantom{$_{-0.0}^{+0.0}$} \\
\hline
PSR~J1705$-$3950 \\
S0 &   $605.4_{-5.9}^{+8.5}$  &  $3.7_{-1.6}^{+2.5}$  &  $5.6_{-2.3}^{+2.2}$  \\ 
S1 &   $588.8_{-6.6}^{+5.7}$  &  $18.5_{-3.1}^{+4.4}$  &  $23.0_{-4.1}^{+4.2}$  \\ 
S2 &   $577.9_{-4.9}^{+6.0}$  &  $11.6_{-3.6}^{+3.5}$  &  $30.4_{-4.4}^{+4.8}$  \\ 
Data  & 577.1\phantom{$_{-0.0}^{+0.0}$} & 11.5\phantom{$_{-0.0}^{+0.0}$} & 32.8\phantom{$_{-0.0}^{+0.0}$} \\
\hline
PSR~J1825$-$1446 \\
S0 &   $668.8_{-7.9}^{+6.7}$  &  $6.1_{-1.5}^{+2.6}$ & -- \\ 
S1 &   $655.2_{-5.9}^{+6.8}$  &  $14.8_{-3.8}^{+6.1}$ & --  \\ 
S2 &   $655.2_{-5.7}^{+6.8}$  &  $14.8_{-3.8}^{+6.1}$ & --  \\ 
Data  & 653.2\phantom{$_{-0.0}^{+0.0}$} & 14.3\phantom{$_{-0.0}^{+0.0}$} & -- \\

\end{tabular}
\end{table}

\subsection{PSR~B1828$-$11}

\begin{figure}
\includegraphics[angle=0,width=0.45\textwidth]{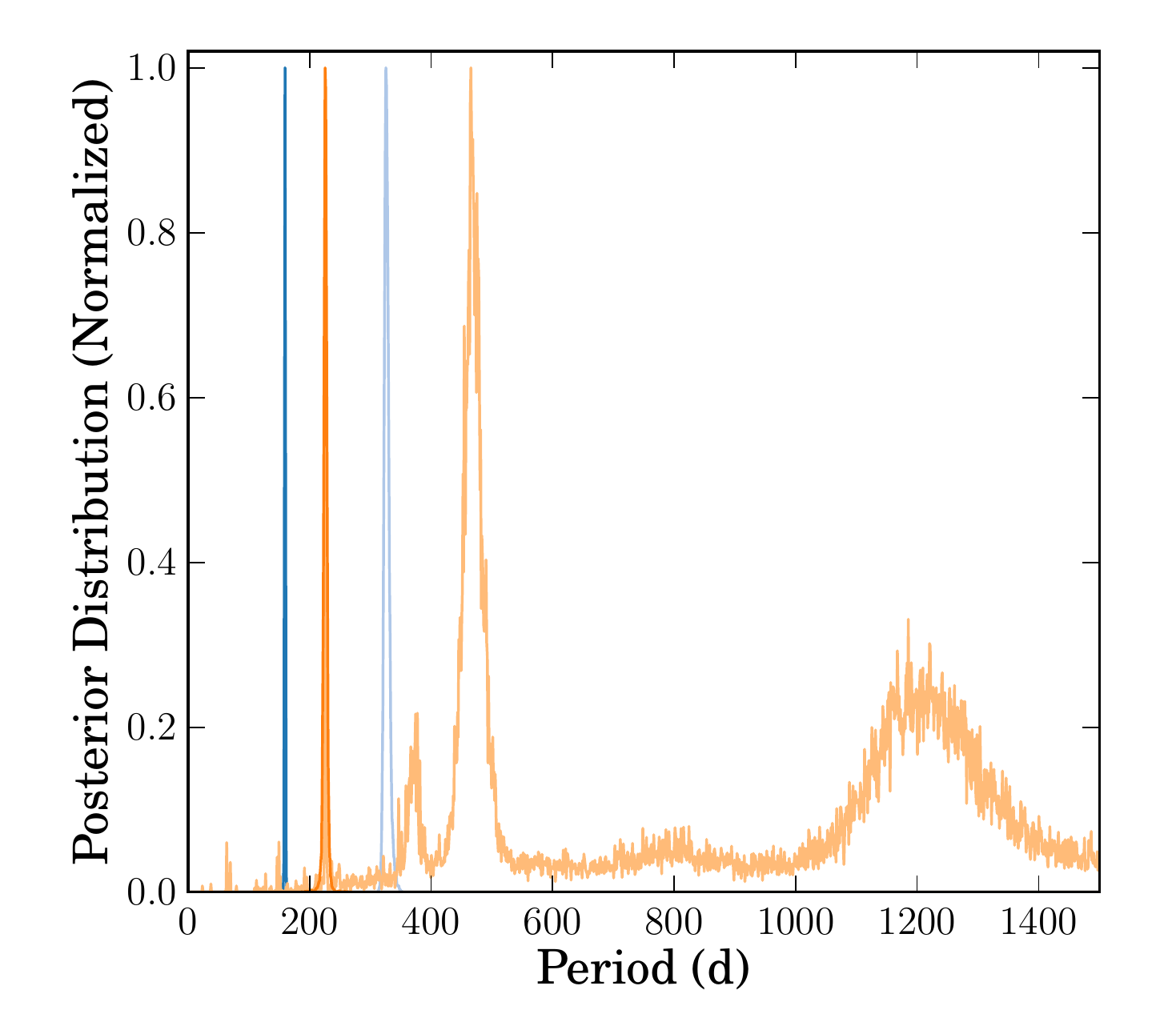}
\caption{\label{fig:j1705_j1830_comp}A comparison of the posterior
probability density distributions for two periodicities in \psra{} and
\psrf{}, all normalized to peak amplitude.  The dark and light blue
lines indicate the two observed periods of \psra{} and imply
well-constrained detections.  The short, 225-day period of
PSR~B1828$-$11 is likewise well-constrained (dark orange), but the
addition of a second period results in many possible solutions of
similar quality (light orange), including a dominant $\sim$450-day
peak.}
\end{figure}

PSR~B1828$-$11 has the most ``well-known'' modulation in our sample,
but it offers an interesting contrast with the other members.  It
appears to have at least one relatively high-quality modulation of
about 225 days, and we also note that this is the dominant time-scale
at which the pulse profile varies, c.f. Figure 5 of \citet{Lyne10}.
It has additional quasi-periodic oscillations with time-scales of
$\sim$\,500 and $\sim$\,1000 days \citep{Stairs00}, all of which have
been interpreted as the ``fundamental'' time-scale by various authors.
However, none of these longer time-scales appears stable.  In Figure
\ref{fig:j1705_j1830_comp}, we show the posterior distributions for a
two-sinusoid model for \psra{} and \psrf{}.  While both the
fundamental and harmonic sinusoid of \psra{} have well-defined values,
only the fundamental of PSR~B1828$-$11 is well defined.  With the
addition of a second sinusoid, solutions at 370, 464, and
790, and 1200 days are of similar quality.

Moreover, while for the seven new pulsars the timing noise that
remains after modelling one or two sinusoids is consistent with a
power law spectrum, even after modelling four sinusoids, the power
spectrum for PSR~B1828$-11$ is inconsistent with a simple
power law, manifesting particularly as best-fit values of
$f_c\sim1$\,yr.  Thus, while PSR~B1828$-11$ clearly supports strong
modulation, it is both more complex and less regular than the seven
candidates. 

\subsection{PSR~B0740$-$28}

Several other pulsars in our parent sample have published analyses
indicating quasi-periodic profile and/or timing modulation.  However,
these examples seem to lack the coherence of those presented here and
did not pass our selection criteria.  E.g., PSR~B0740$-$28 shows
marked, rapid profile and spin-down variations with a time-scale of
about 50 days \citep{Lyne10}, but these oscillations are irregular.
Interestingly, \citet{Keith13} found that correlation between the
spin-down properties and profile increased dramatically following a
glitch, and \citet{Brook15} found similar behaviour
using complementary methods.

\subsection{Summary}

To summarise, the candidates we have identified have three key
features: (1) the modulations are of high quality; (2) some show
significant modulation at a harmonic, but with amplitudes about ten
times smaller than the fundamental; and (3) after removing the
sinusoidal modulation, the remaining the timing noise has a spectrum
well described by a simple power law.  The high quality factor
distinguishes these objects from the ``state switching'' pulsars
presented by \citet{Lyne10} and \citet{Hobbs10b}.  A further key
feature of about one third of this ``state switching'' sample is the
presence of pulse profile variation, and we proceed to search for such
variation in our sample.

\section{Search for Pulse Profile Variations}
\label{sec:profile}

If the TOA modulations presented here represent a high-$\dot{E}$
continuation of the magnetospheric switching observed by
\citet{Lyne10}, there we might be evidence for pulse profile
variation.  Unfortunately, most of our candidates are relatively
radio-faint, with typical signal-to-noise ratio of 10--20 in our short (typically
2--3\,minute) observations.  Several pulsars also suffer strong
scattering by turbulent electrons in the interstellar medium, and the
multi-path propagation dilutes intrinsic profile changes.

To mitigate these limitations, we have adopted the following approach
as the most sensitive to pulse profile variations correlated with TOA
modulations: if the periodic behaviour is due to magnetospheric
switching between $\dot{\nu}$ states, then the ``zero crossing'' of
the residual time series indicates a switch.  Timing noise complicates
this prescription, and we have instead taken the best-fitting
realization of the sinusoidal signals discussed above as indicative of
possible state changes and co-added the profiles from the intervals
with positive and negative residuals.  Because the received flux
density varies due to scintillation in the interstellar medium, before
co-addition we normalized each profile by its measured flux density.

\begin{figure*}
\includegraphics[angle=0,width=0.95\textwidth]{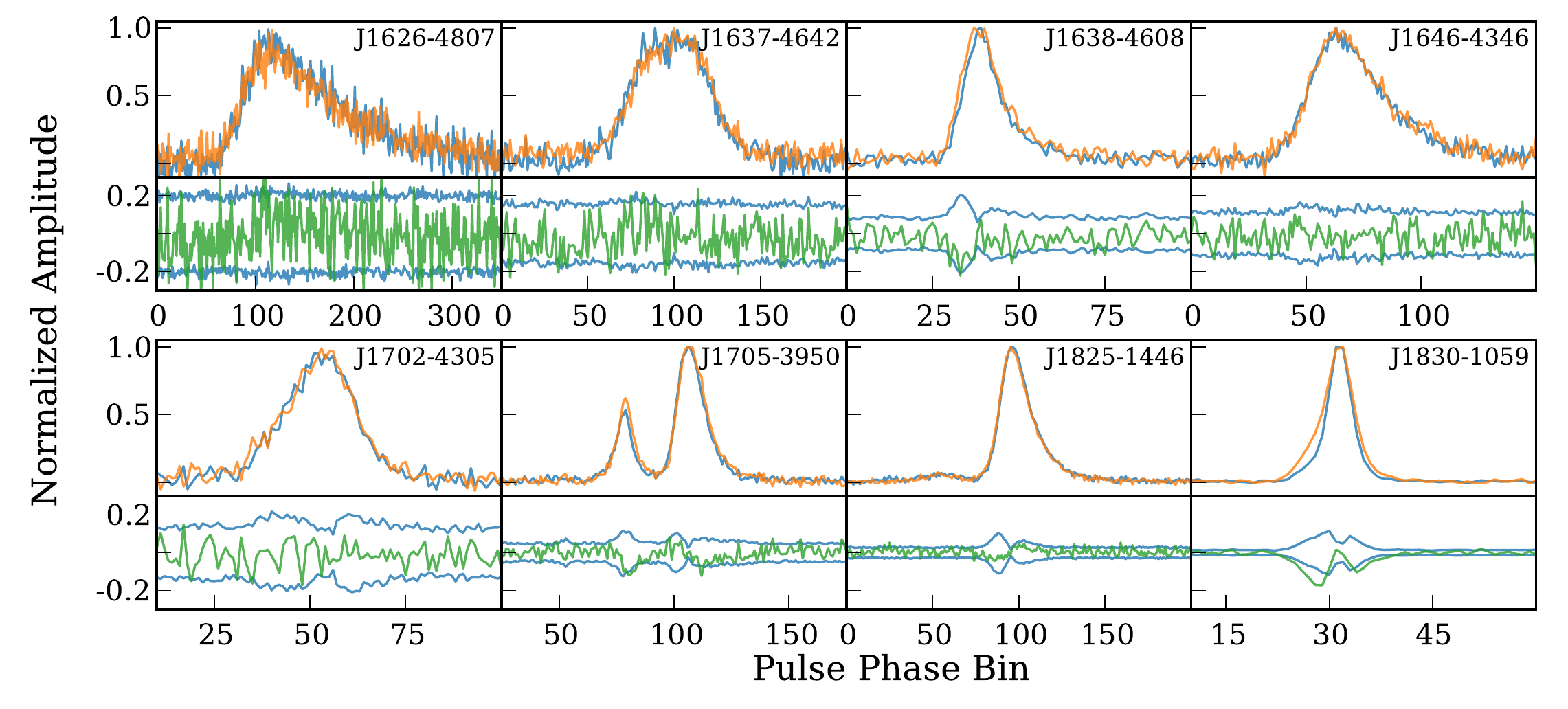}
\caption{\label{fig:prof_var}Analysis of profile variation.  The full
profiles have 512 phase bins and are zoomed to the indicated phase
bins.  The orange and blue traces in the upper panel show the co-added
profiles obtained during periods of positive and negative timing
residual, normalized by flux and aligned by TOA.  The green lower
panel trace shows the difference.  The light blue trace shows the
central 90\% of differences resulting from a random shuffling of the
individual profiles among groups, giving an estimate of the
statistical significance of the observed differences.}
\end{figure*}

The resulting co-added profiles appear in Figure \ref{fig:prof_var}.
The lower panels show the difference between the two ``states''.  To
determine whether any differences are statistically significant, we
repeated the exercise 100 times but randomly assigned profiles to one
of the two states and have plotted the envelope resulting from these
simulations in the lower panel.  Four pulsars are dominated by white
noise and/or scattering and show no evidence of correlated profile
variation: J1626$-$4807, J1637$-$4642, J1646$-$4346, and J1702$-$4306.
Three pulsars, J1638$-$4608, J1705$-$3950, and J1825$-$1446, show
differences significant relative to the radiometer noise level.
However, the differences can be accounted for by chance, suggesting
the pulse profile variation is due to jitter rather than correlated
with the TOA modulation.  Finally, PSR~B1828$-$11 shows a clear
correlation between the pulse profile state and the TOA modulation,
with a narrower profile occurring when the TOA residuals are negative.

The observed profile differences in PSR~B1828$-$11 are significant
according to both of our tests but are substantially smaller than
those observed by \citet{Stairs00}.  This is due to our adoption of a
single sinusoid model, which as discussed above does not capture the
full modulation behaviour of the pulsar.  We expect it to work well
for the other candidates whose modulation is dominated by the
fundamental frequency.

The profile differences in J1705$-$3950, and J1825$-$1446 hint at real
correlated profile changes and motivate a dedicated campaign to obtain
deep observations at higher frequencies where the effects of
scattering are reduced.  We conclude, however, that aside from the
known profile variation of PSR~B1828$-$11, there is no significant
profile variation in the new candidates.  \citet{Brook15} reach a
similar conclusion using a Gaussian regression method that smoothly
interpolates both profile shapes and spin-down variations between
observations.  This ``model independent'' approach avoids the need to
classify observations into particular states and offers complementary
verification of our results.  The lack of pulse profile variation,
together with the sinusoidal nature of the TOA modulation, suggests
our candidates are indeed distinct from previous examples.  Below, we
consider mechanisms that may be responsible for the observed
properties.

\section{Interpretation of Periodic Signals}
\label{sec:discussion}

We have established that seven pulsars in our sample show stable,
sinusoidal modulation and have no strong evidence for pulse profile
variation.  The regularity of these modulations requires a stable
underlying clock, and the $>$100-day time-scales involved are far
longer than any dynamic process expected from the neutron star
magnetosphere.  We thus expect the clock to be either external to the
magnetosphere or associated with the neutron star itself, and we
discuss a scenario for each case below.  First, planetary companions
naturally furnish sinusoidal modulation through reflex motion and
should not perturb the pulse profile.  Second, we consider the
``biased magnetospheric switching'' proposed by \citet{Jones12}, in
which neutron star precession drives regular magnetospheric state
switching.  The resulting current reconfiguration modulates the
external torque, and hence the pulse TOAs, and the pulse emission
profile.

\subsection{Planets}

\begin{table}
\centering
\caption{\label{tab:planets}Derived parameters for periodic signals;
Planetary values assume a circular orbit about a 1.4\,$M_{\sun}$ neutron star.  $\delta\dot{\nu}$ is computed as $(2\,a/P)(2\pi/P_{b})^2$.}
\begin{tabular}{lrrrr}
\hline
 & $P_b$ & $m_c\sin i$ & $a$  & $\delta\dot{\nu}/\dot{\nu}$ \\
 & yr & $M_{\earth}$ & AU & \% \\
\hline
J1626$-$4807 & 0.76 & 3.03 &  0.95 & 0.68 \\
J1637$-$4642 & 0.52 & 0.64 & 0.65 & 0.04 \\
J1638$-$4608 & 1.29 & 6.17 &  1.61 & 0.20\\
        & 0.65 & 0.91 &  0.81 & \\
 J1646$-$4346 & 0.33 & 4.54 & 0.41 & 0.38 \\
J1702$-$4306 & 1.07 & 2.41 &  1.34 & 0.36 \\
        & 0.54 & 0.55 &  0.68 & \\
J1705$-$3950 & 0.89 & 2.44 &  1.12 & 0.14 \\
        & 0.43 & 1.01 &  0.55 & \\
J1825$-$1446 & 0.34 & 0.62 & 0.42 & 0.31 \\
\end{tabular}
\end{table}

If interpreted as planetary companions, the TOA modulations correspond
to earth-mass planets which reside within two AU of the neutron star
(Table \ref{tab:planets}).  As we discussed in \citet{Kerr15}, 
forming such massive planets within $10^5$\,yr, the typical age
of these pulsars, is challenging but not impossible, provided a disk
of sufficient angular momentum is able to deposit copious dust in a
relatively narrow radial range \citep{Hansen09}.  The radii are large
enough that the effects of both tidal disruption and
evaporation/heating in the pulsar wind are negligible on the growth of
rocky bodies.

However, there are several arguments against a planetary
interpretation.  First, why have none been detected about older
pulsars?  Planets of these masses would be easily identifiable in such
timing residuals, particularly since the level of stochastic timing
noise is also reduced.  Their absence in samples such as those of
\citet{Hobbs06} suggests our candidates are not planets.  On the other
hand, planetary systems may be selected against in samples of
primarily older pulsars.  The orbital planes of binary systems will
preferentially be more edge-on than face-on.  Thus, if the neutron
spin axis is aligned with the plane of its accretion/protoplanetary
disk, alignment of the magnetic field with the spin axis on
time-scales of $<$10$^7$\,yr would decrease the rate of planetary
systems with visible pulsar beams.  However, alignment is a slow
process, if it happens at all \citep[e.g.][]{Tauris98}, so we conclude
that the lack of planet detections in other samples argues against
such an interpretation for the majority of our candidates.

Secondly, at least one, and possibly three, of our candidates boast a
significant harmonic modulation, implying two planets in 2:1
mean-motion resonances.  \citet{Beauge03} have studied this resonance
for a range of parameters, including the mass ratios $m_2/m_1>1$ we
find in all of our two-sinusoid systems.  The stable resonance
condition requires aligned apsides and, generally, a large
eccentricity for the inner (less massive) planet.  The epoch of
periastron measurements for the pulsars with significant second
harmonics are generally not consistent with aligned apsides.
Intriguingly, though, the strongest second harmonic candidate,
PSR~J1705$-$3950, is consistent with an aligned apsides orbit, and its
implied planet masses are some of the lowest in the sample.  If any of
these signals correspond to planetary companions, we advance this
pulsar as the most likely candidate.

On the other hand, the presence of the second harmonic could be
explained with a single planet in a moderately eccentric orbit.  We
have performed fits of the timing modulation of \psra{} and \psrb{}
with a single, eccentric planet.  While we find that the eccentricity
is generally significant, the two-sinusoid models are statistically
preferred, and the measured eccentricity values of 0.4 (\psra{}) and
0.2 (\psrb{}) are somewhat larger than expected from oligarchical
formation scenarios \citep{Hansen09}.


\subsection{Precession}

Precession---and, specifically, free precession---coupled as it is to
the large neutron star moment of inertia, makes an attractive
candidate for the clock underpinning high-quality modulations.  For
a biaxial rotator precessing with a small ``wobble angle'', the
precession frequency $\nu_p$ is simply related to the spin frequency
$\nu_s$ by the Eulerian relation
\begin{equation}
\label{eq:prec_freq}
\frac{\nu_p}{\nu_s} = \frac{P_s}{P_p} = \frac{\Delta I_p}{I_p} \equiv \epsilon
\end{equation}
where $\Delta I_p$ is the difference in moment of inertia between the
symmetry axis and either of the other two axes and $I_p$ is the
characteristic moment of inertia of the \textit{precessing body}.  A
pedagogical derivation of this result can be found in \citet{Jones01},
and more complete treatments in \citet{Cutler03}, \citet{Sedrakian99},
and \citet{Wasserman03}.  A key point is that $I_p$ includes any parts
of the neutron star tightly coupled (on time-scales $\leq P_s$) to the
precessing crust.

Although this result is often presented in the literature, there is
widespread disagreement about both $\Delta I_p$ and $I_p$, i.e. which
parts of the neutron star are coupled to precessional motion, as well
on damping time-scales for precession.  Further complications include
the r\^{o}le of external torque applied by the magnetosphere; although
it generally modifies the precession dynamics only slightly, it is
predominantly the time-varying external torque that produces TOA
modulation.  We outline these arguments and their observational
implications below.

\subsubsection{The Precession Frequency}

For pulsars spinning at 1\,Hz to precess with $\nu_p=1/\mathrm{yr}$,
Equation \ref{eq:prec_freq} implies an \textit{ellipticity}
$\epsilon = 3\times10^{-8}$, which in turn requires the correct ratio
between neutron star deformation $\Delta I_p$ and total moment of
inertia coupled to the precession $I_p$.

Several mechanisms have been proposed for establishing $\Delta I_p$.
E.g., magnetic stresses associated with the internal dipole field can
provide an ellipticity of order the ratio of magnetic field energy to
gravitational field energy $\epsilon\approx10^{-12}$
\citep[e.g.][]{Zanazzi15} with the ellipticity aligned with the
magnetic moment.  Together with intrinsic ellipticities, such
deformations may be relevant for the claimed secular evolution of the
magnetic field inclination of the Crab pulsar \citep{Lyne13}.  For
large stresses, the neutron star is inherently triaxial, and
interestingly, such models naturally produce relatively large
modulation power in harmonics, as we observe in PSR~B1828$-$11 and
some members of our sample \citep{Wasserman03}.  Such large stresses, about
two orders of magnitude above the dipole contribution, could result
from a Type II superconducting core or strong toroidal fields.  As we
note below, though, a Type II superconducting core may inhibit low
frequency precession through pinning of core neutron superfluid
vortices to proton superfluid flux tubes.

A relatively natural mechanism for achieving the ellipticities
required for $P_p=1$\,yr is the strained crust.  In this scenario, the
crust solidifies when the neutron star is spinning rapidly, freezing
in a zero strain oblateness proportional to its large spin frequency,
$\epsilon_0\propto\nu_0^2$.  As the neutron star spins down and
rotational support is removed, the crustal bulge decreases, but
Coulomb forces act to maintain a deformation larger than that
supported by the current centrifugal force.  This deformation,
which depends on the ratio of the Coulomb and gravitational forces, is
parameterized by the \textit{rigidity parameter} $b\equiv\Delta
I_p/\Delta I(\nu)$, with $\Delta I (\nu)$ the equatorial bulge
\citep{Jones01,Cutler03}, and is estimated to be about
$2\times10^{-7}$ \citep{Cutler03}.  If the crust is relaxed at the
current spin rate, such a value produces ellipticity of order
$3\times10^{-11}$, while a strained crust with reference spin $\nu_0$
will increase the ellipticity by $(\nu_0/\nu_s)^2$, more than two
orders of magnitude if the neutron star is born spinning at tens of
Hz.  More precisely, \citet{Cutler03} find
\begin{equation}
\label{eq:cutler}
\frac{\Delta I_p}{I_p} \approx 9.7\times10^{-9}
\left(\frac{\nu_0}{40\,\mathrm{Hz}}\right)^2\frac{I_p/I_*}{10^{-2}}\left(\frac{M}{1.4\,M_{\sun}}\right)^2\left(\frac{R}{12\,\mathrm{km}}\right)^4,
\end{equation}
where as before $I_p$ is the moment of inertia involved in precession,
and $I_*$ is the total moment of the neutron star.  As can be seen
from the strong dependence on $R$, this ellipticity can
be varied by a factor of two simply by stiffening/softening the
neutron star equation of state.

\subsubsection{Component Coupling}

A neutron star is substantially more complicated than the picture
presented above.  The core---containing about 90\% of the moment of
inertia---is thought to comprise primarily a neutron superfluid
and a few per cent by mass protons (also superfluid) and
electrons (degenerate).  Some of the neutron superfluid is thought to
penetrate the crust and its vortices pin on the crustal lattice; the
spontaneous or triggered release of these pinned vortices is a
plausible mechanism for the large glitches observed in young pulsars
\citep[e.g.][]{Anderson75,Alpar88}.

If some portion of superfluid is pinned to the crust, it dramatically
increases the effective ellipticity (and precession frequency) as
first noted by \citet{Shaham77}.  Specifically, for small wobble
angles, $\nu_p/\nu_s\approx I_{pin}/I_{p}$, where the numerator is the
moment of inertia of the pinned superfluid and the denominator the
moment of inertia of all other precessing components.  Assuming $I_p$
is primarily due to the crust and coupled charged protons in the core,
the right-hand side of this relation is unity, ruling out long-period
precession.

However, \citet{Link02} have calculated that Magnus forces due to
precession with modest wobble angles (3\degr) as proposed for
PSR~B1828$-$11 exceed the force keeping vortices
pinned on crustal nuclei.  On the other hand, precession at small
wobble angles (0.1\degr) could allow substantial pinning.
Consequently, large wobble angle precession and large amplitude
glitches should be mutually exclusive phenomena, and small-amplitude
precession at low frequencies may be prevented by superfluid pinning.


Equally critical is the core / crust coupling.
\citet{Alpar88} calculated the coupling due to the
magnetisation of superfluid vortices via entrainment of protons and
electrons and the subsequent scattering of charged particles.  This
effect couples the core to the crust on time-scales of
400--10,000\,P$_s$.  This coupling is weak in terms of precession,
and the core moment of inertia is effectively decoupled from the
precession.  The coupling will weakly damp precession on time-scales
of 400--10,000\,P$_p$ \citep{Sedrakian99}.  Although we will not
discuss it here, we note that instabilities arising from the velocity
differential between the two superfluids may complicate this picture
\citep[][]{Glampedakis09}.

\citet{Link03} notes that interaction between neutron superfluid
vortices and magnetic flux tubes in the proton superfluid in the core
\citep[see also][]{Chau92}
can either (1) lock the core to the crust (short coupling time) or (2)
dissipate precessional motion on time-scales of hours via kelvon
excitation.  He concludes $P_p=$1\,yr precession implies that the
core proton superfluid is a Type 1 superconductor, effectively
separating vortices and flux tubes.  However, the details of these
calculations depend sensitively on the poorly known distribution of
the constituents of the core, e.g. strong toroidal fields
\citep[e.g.][]{Sidery09}.  Indeed, \citet{Haskell13} argue that such
pinning would also prohibit the large glitches observed in
PSR~J0835$-$4510 (Vela).

\subsubsection{External Torques}

Until now, we have neglected external torques in our discussion.
However, they are important in establishing the observational
implications of precession.  Generally, these torques trivially
modify the precessional motion of interest, but the spin-down rate of
the neutron star depends on the magnetic inclination $\alpha$, and
precession gives $\alpha$ dependency on time.  Indeed, even in the
absence of precession, external torques tend to force
$\alpha\rightarrow0$, but only on the pulsar spin-down time-scale
\citep{Melatos00}.  Most calculations of, e.g., the time-dependent
spin-down rate under precessional motion are in the vacuum limit
\citep{Link01,Jones01,Wasserman03}.  Their application to
PSR~B1828$-$11 results in extremely stringent constraints on the
magnetic inclination, viz.  that it be orthogonal to the spin axis to
within 1\degr, a prediction at odds with the lack of an observed
interpulse from the opposite magnetic pole.  More recent calculations
make use of numerical modelling of force-free magnetospheres, with
substantially altered torques on the neutron star surface
\citep{Arzamasskiy15} that allow for broader range of $\alpha$ in
describing PSR~B1828$-$11.  Interestingly, dominant magnetic stresses
allow an even larger range of $\alpha$ \citep{Wasserman03,Akgun06}.

\subsubsection{Observations}

Despite the large body of work on neutron star precession, whose
surface we have only scratched here, the verdict still seems to be out
on whether or not (1) it can proceed at all; (2) it can operate on
annual time-scales; (3) it can persist as a high-quality-factor
signature in pulse times of arrival.  We therefore proceed in an
empirical fashion and study the implications of the observed
modulations for a precession interpretation.

\begin{figure}
\includegraphics[angle=0,width=0.45\textwidth]{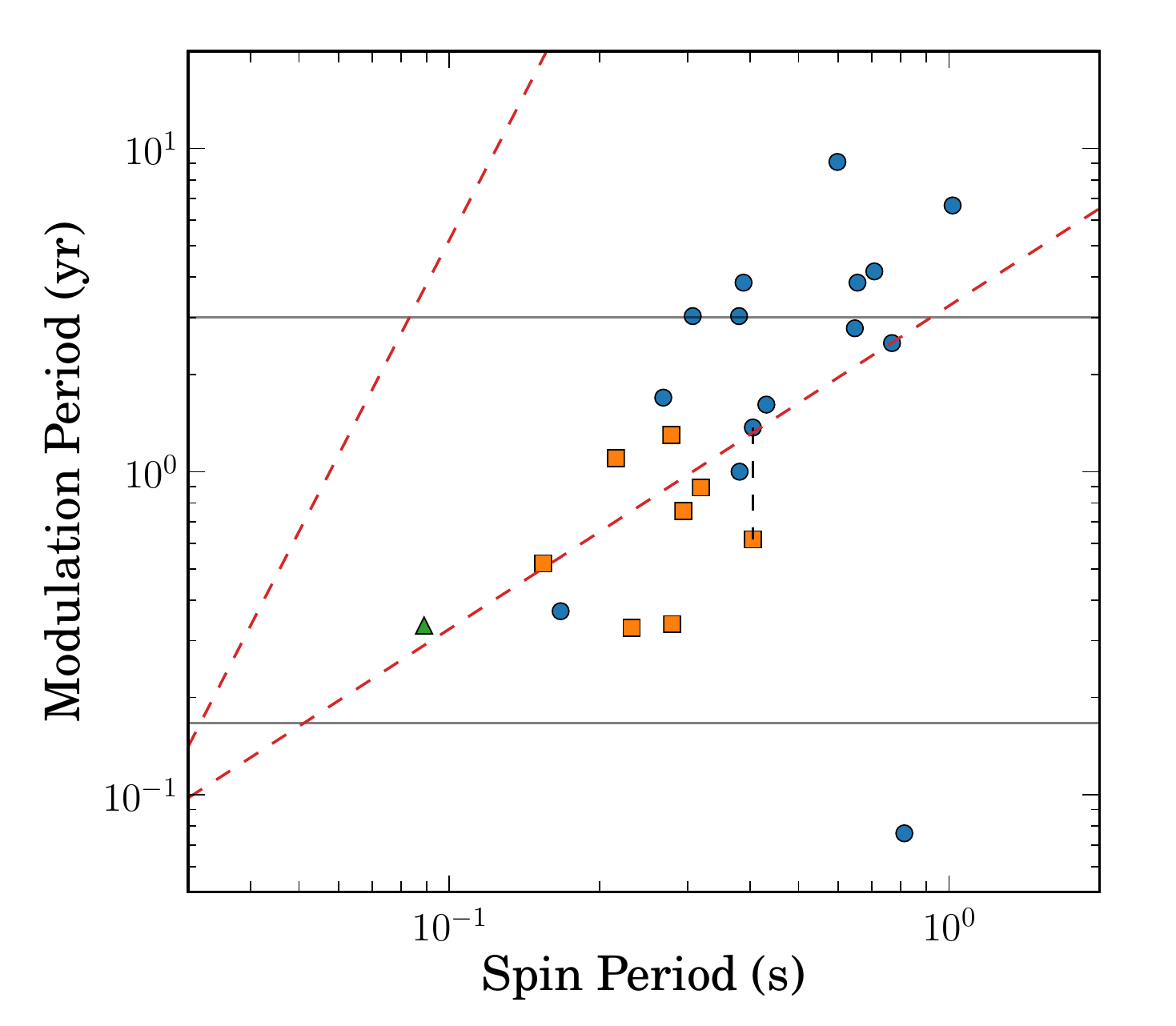}
\caption{\label{fig:mod_vs_period}The observed modulation period for
our sample (orange squares) and the 15 pulsars from the \citet{Lyne10}
sample as characterised by \citet[][blue circles]{Jones12}.  The green triangle shows
the 122\,d period inferred by \citet{Durant13} from the helical jet of
PSR~J0835$-$4510.  The solid, horizontal
lines indicate, approximately, the range of modulation periods to
which our analysis is sensitive.  The dashed lines represent the
expected modulation period for 1.4\,$M_{\sun}$, 12\,km neutron stars
with a frozen in strain corresponding to a reference spin of 40\,Hz
(see Equation \ref{eq:cutler}; linear dependence on spin period
and shorter modulation periods) and those for similar neutron
stars with relaxed crusts ($\propto P_s^3$, and longer modulation
period).  We have shown PSR~B1828$-$11 with our measured period (225
days) as well as that calculated by \citet{Jones12}, about twice as
long.  The outlier with modulation period $<$0.1\,yr is
PSR~B1931$+$24.}
\end{figure}

In Figure \ref{fig:mod_vs_period}, we have plotted the measured
modulation periods from our sample (orange) along with the values
measured by \citet{Jones12} from the sample of \citet{Lyne10} as a
function of pulsar spin period.  If the modulation is due to
precession of a neutron star with a strained crust, $P_p\propto P_s$,
and a reference spin of 40\,Hz yields a strain level that reproduces
the observed modulation period within an order of magnitude.  However,
the observed dependence is somewhat steeper, with a best-fit relation
$P_m/\mathrm{yr}=7.1\,P_s^{1.5}$.  On the other hand, if the neutron
star crusts are fully relaxed at the current spin period, $P_p\propto
P_s^3$ and, due to the reduced ellipticity, predictions for $P_p$ are
much longer.  Finally, we note that some of the period measurements
from \citet{Jones12} may be biased to larger values: these
oscillations are of lower quality factor and some cycles may be
``missing'', in which case the measurement procedure gives a result
longer than the true modulation period.  Correcting this bias would
alleviate some of the tension with the strained crust picture.

The obvious outlier in Figure \ref{fig:mod_vs_period} is the
intermittent pulsar PSR~B1931$+$24 \citep{Kramer06}, which we have
included for comparison.  None of the mechanisms described above are
plausible if its intermittency is interpreted as precession.

We have also included the recent detection of helical motion in the
X-ray jet of Vela \citep{Durant13} with an apparent period of 122\,d,
although shorter periods (as aliases) are allowed.  If the corkscrew
motion is due to pulsar precession, the measured value is in general
agreement with the observed trend.

\begin{figure}
\includegraphics[angle=0,width=0.45\textwidth]{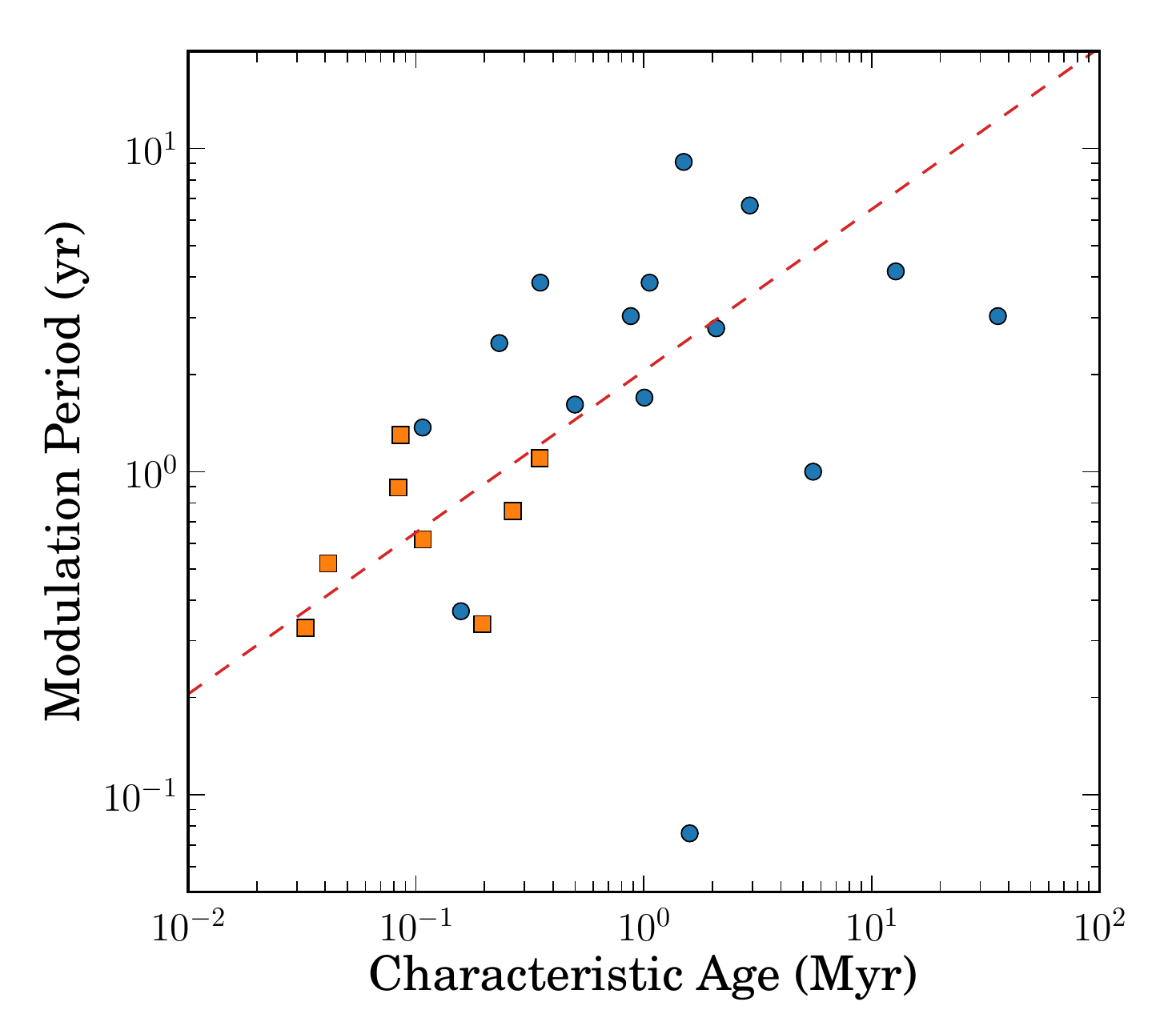}
\caption{\label{fig:mod_vs_age}As Figure \ref{fig:mod_vs_period} but
with characteristic age as abscissa.  The dashed line gives the
prediction of Equation \ref{eq:mag_stress}.}
\end{figure}

Next, we consider a scenario in which magnetic stresses dominate the
ellipticity.  Dipole stresses are generally too small to provide the
observed periods, but for completeness we note they predict
$P_p\propto1/\dot{P_s}$, i.e. no dependence on $P_s$, in poor
agreement with the observations.  For a superconducting core
\citep{Wasserman03}, \citet{Jones12} gives the relation as
\begin{equation}
\label{eq:mag_stress}
\frac{P_p}{1\,\mathrm{yr}}=2.0\left(\frac{\tau_{sd}}{10^6\,\mathrm{yr}}\right)^{1/2}\left(\frac{2\times10^{15}\,\mathrm{G}}{H_c}\right)\left(\frac{I_p}{I_*}\right)
\end{equation}
with $\tau_{sd}\equiv P_s/2\dot{P_s}$
the characteristic pulsar age and $H_c$ the critical field due to the
superconducting core.  Accordingly, we have plotted the modulation
periods against characteristic age in Figure \ref{fig:mod_vs_age}
along with the trend of Equation \ref{eq:mag_stress}.  The relation
describes both the magnitude and trend of the data reasonably well.
But, we note that the predicted modulation periods require the entire
neutron star to participate in precession ($I_p=I_*$).  If, instead,
only the crust participates ($I_p\approx0.01I_*$), the precession
periods are too rapid to fit the observations.

\begin{figure}
\includegraphics[angle=0,width=0.45\textwidth]{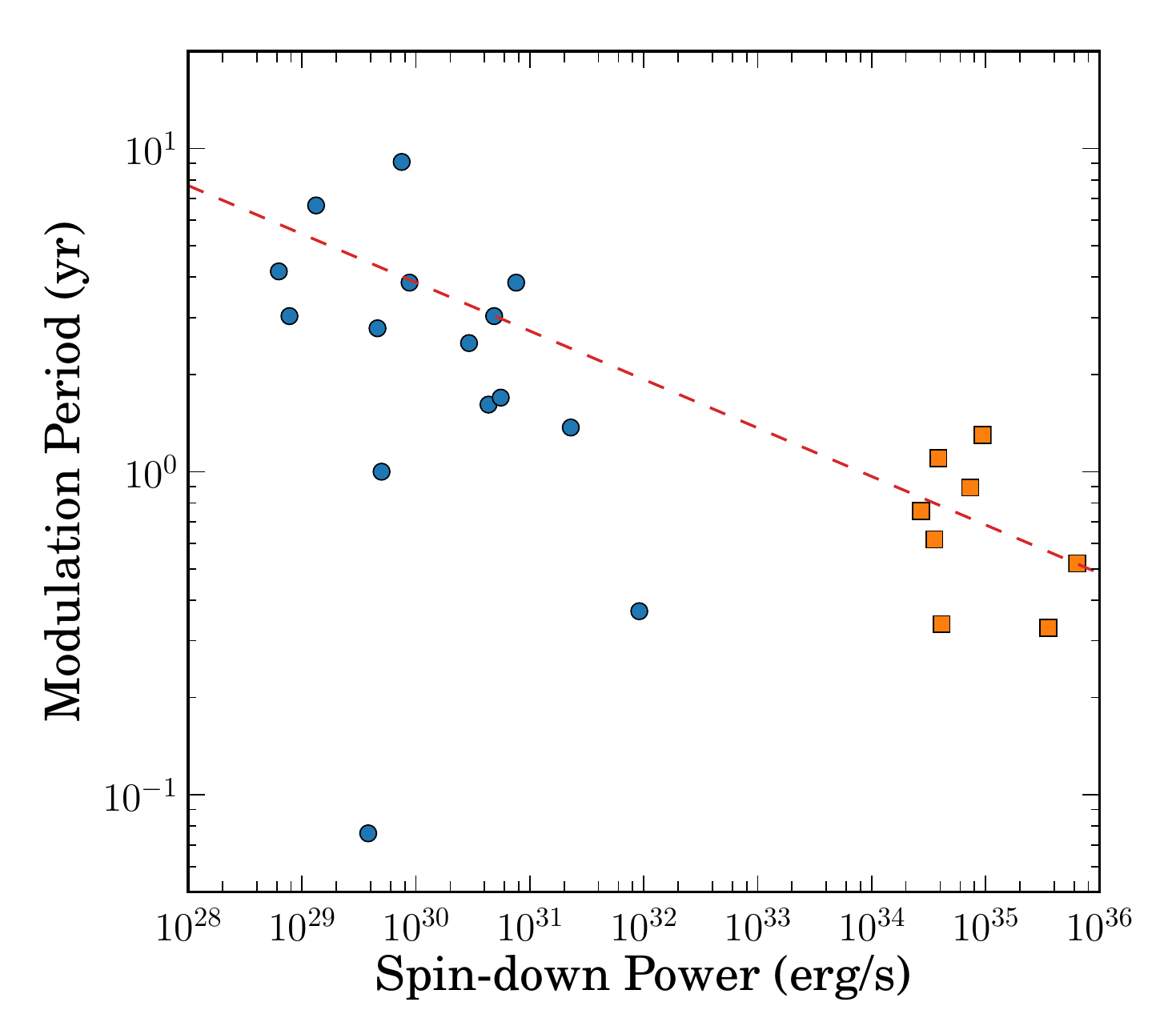}
\caption{\label{fig:mod_vs_edot}As Figure \ref{fig:mod_vs_period} but
with spin-down power as abscissa.  The line is simply to guide the eye
and is $\propto\dot{E}^{-0.15}.$}
\end{figure}

In Figure \ref{fig:mod_vs_edot}, we show the dependence of
the observed modulation period on spin-down power ($\dot{E}\propto
\dot{P_s}/P_s^3$).  The scatter is relatively narrow, in keeping with
our finding that the data depend maximally on $P_s$ and little on
$\dot{P_s}$, i.e. $P_m/\mathrm{yr} =8.6\,P_s^{1.35}\dot{P_s}^{-0.15}$.
We also see an interesting gap in parameter space coverage, motivating
a timing program targetting pulsars in the range
$10^{31}<\dot{E}/\mathrm{erg\,s^{-1}}<10^{34}$.

We conclude this section by briefly considering millisecond pulsars
(MSPs).  The timing precision can easily reach $<$1\,$\mu$s,
particularly for pulsars monitored by timing arrays
\citep[e.g.][]{Manchester13}.  Despite this sensitivity, no similar
harmonic oscillations have been observed.  This result is not
surprising, however, if any of the precession mechanisms proposed
above are in play.  During the spin-up process, neuton star crusts are
likely to break under the substantial centrifugal force and
distribution of accreted material, so the crusts of current MSPs are
likely to be relaxed, and Equation \ref{eq:cutler} indicates
ellipticies of about $10^{-6}$ in this case, with precession periods
of about an hour.  In the case of magnetic stresses, the much weaker
dipole magnetic fields would push the precession period higher, even
when allowing for a superconducting core.  Equation
\ref{eq:mag_stress} with a typical MSP age $\tau_{sd}=10^9$\,yr
suggests a precession period of about 60 years, though for crust-only
precession this drops to 1--2 years.  All cases would require some
continual mechanism to excite the precession in order for it to be
observable.

\section{Discussion}

As we have discussed above, we cannot rule out planets as the source
of the nearly sinusoidal TOA modulations we observe and perhaps one of
our systems harbours \textit{bona fide} planets.  Our preferred clock,
however, is neutron star precession, but the extent to which it can
operate in real neutron stars is unclear.  Regardless of the actual
mechanism, we can ask: is the same mechanism responsible for both the
longer period, lower quality oscillations of \citet{Lyne10} and the
shorter period, higher quality modulations of our sample?  To address
this, we amplify on the idea of \citet{Jones12}, viz. that a pulsar
may lie on the cusp of two different magnetospheric configurations,
and that the precessional phase alters the probability that the pulsar
may switch between them.  \citet{Cordes13} has analyzed a similar
scenario with forced Markov processes to model state-changing pulsars.
In both situations, the altered magnetosphere acts as a ``lever arm''
to enhance the observational signature of potentially minor
underlying changes.

Can such a mechanism unify the two samples?  We present a qualitative
argument that it can.  The most dramatic changes in spin-down rate, of
order unity, are associated with intermittent pulsars
\citep{Kramer06,Camilo12,Lorimer12}.  The magnetospheric currents are
clearly dramatically reconfigured during the state change that turns
the pulsar off.  Large changes in the plasma density of the
magnetosphere of PSR~B0943$+$10 \citep{Hermsen13} happen rapidly when
the pulsar changes between its radio-bright and radio-faint states.
These pulsars tend to be at least 1\,Myr old, suggesting that the
stable states of old pulsar magnetospheres are separated by large
energy barriers.  In contrast, the modulated pulsars of \citet{Lyne10}
show much more modest changes in $\dot{\nu}$, typically 1\%, and the
pulse profile variations can be quite subtle.  For our sample, the
fractional $\dot{\nu}$ changes are even smaller (Table
\ref{tab:planets}), all well less than 1\%, and as discussed in
\S\ref{sec:profile}, there is no evidence for profile variation.

Thus, we posit that pulsars have one or more metastable magnetospheric
states \textit{whose properties change with age}.  For young pulsars,
these states are finely separated in activation energy, and may even
form a continuum.  As the pulsar ages, its magnetosphere evolves
discrete states more widely discrepant in currents, beam patterns, and
spin-down torques; in some states the current configurations may limit
particle acceleration processes.  Thus, in young pulsars, very small
amplitude (wobble angle) precession could smoothly modulate the
current structure in the magnetosphere, with the latter enhancing the
change in spin-down rate from that expected simply from changing the
magnetic inclination \citep[c.f.][]{Arzamasskiy15}.  With increasing
age, the transition between states becomes more stochastic, the higher
activation energy preventing the magnetosphere from smoothly following
the precession.  The quality factor of the modulations drops, and
profile variation becomes evident, e.g. as in the $\sim$600\,d cycle
of PSR~B0916$+$06 \citep{Perera15}.  In some cycles, a state change
may not occur at all, perhaps explaining ``missing'' $\dot{\nu}$
cycles like those visible in the time series of PSR~B1642$-$03
\citep{Lyne10}.  Extreme versions of this situation may account for
``episodic'' events like the profile change of PSR~J0738$-$4042
\citep{Brook14}, attributed by those authors to an encounter with an
asteroid.  For that event, we argue that the continued profile changes
after the sudden shift may be more easily attribute to an ongoing
process like precession.  Eventually, the precession is no longer able
to drive the star into different states, and whatever pulse modulation
effects occur through modulation of $\alpha$ are too small to observe.
Precessionaly motion itself may likewise be difficult to maintain in
older pulsars if glitches or timing noise are responsible for its
excitation.

Indeed, glitches add two final complicating anecdotes.  First, the
\textit{change} in correlation between spin-down and profile variation
observed by \citet{Keith13b} for PSR~B0740$-$28 suggests that
relatively minor variations in the neutron star rotation can indeed
effect magnetospheric changes, in support of our scenario above.
Second, the TOA modulation we observed in PSR~J1646$-$4346 appears to
be consistent through a large glitch
($\delta\nu=3.7\times10^{-5}$\,Hz, $\delta\nu/\nu=8.5\times10^{-6}$,
$\delta\dot{\nu}/\dot{\nu}=0.61$\%). The substantial amplitude implies
a large amount of the core superfluid was pinned prior to the glitch,
necessitating small wobble angles to preserve the pinning.  Such
substantial superfluid pinning would tend to raise the precession
frequency dramatically.

\section{Summary and Conclusion}

We have detected nearly-sinusoidal TOA modulation in seven members of
a population of 151 energetic pulsars.  Other candidates may lurk
below our sensitivity limit, indicating such modulation is a
relatively common phenomenon.  This prevalence, together with the
regularity of the modulation, seems to demand a dynamic process rooted
in the neutron star, and we suggest precession may provide the
``clock''.  Observational and theoretical difficulties with modest
wobble angles can be resolved by letting very small wobbles drive
magnetospheric state switches.  Due to the strength of the harmonics
in the the TOA modulations, as well as the expectation that the core
should couple tightly to the crust, we propose magnetic stresses
supported by the superconducting core as the most likely source of the
ellipticity required for precession.

\section*{Acknowledgements}

The Parkes radio telescope is part of the Australia Telescope, which
is funded by the Commonwealth Government for operation as a National
Facility managed by CSIRO.  This research has made use of NASA's
Astrophysics Data System, for which we are grateful.

\bibliographystyle{mn2e}


\label{lastpage}

\end{document}